%% file: main.tex
\documentclass[preprint,12pt,authoryear]{elsarticle}

\usepackage{amssymb}
\usepackage{amsmath}

\usepackage{hyperref}
\usepackage{siunitx}

\usepackage{rotating}

\newcommand{\fig}[1]{Figure~\ref{fig:#1}}
\renewcommand{\sec}[1]{Section~\ref{sec:#1}}
\newcommand{\tab}[1]{Table~\ref{tab:#1}}

\journal{Astronomy and Computing}

\begin{document}

\begin{frontmatter}



% \dochead{}

\title{\textsc{Sengi}: a small, fast, interactive viewer for spectral outputs from stellar population synthesis models}


\author{Christopher C. Lovell}
\address{School of Physics, Engineering and Computer Science, University of Hertfordshire, Hatfield, AL10 9AB, UK}
\ead{c.lovell@herts.ac.uk}

\begin{abstract}
We present \textsc{Sengi}, (\href{https://christopherlovell.github.io/sengi}{https://christopherlovell.github.io/sengi}), an online tool for viewing the spectral outputs of stellar population synthesis (SPS) codes.
Typical SPS codes require significant disk space or computing resources to produce spectra for simple stellar populations with arbitrary parameters.
This makes it difficult to present their results in an interactive, web-friendly format.
\textsc{Sengi} uses Non-negative Matrix Factorisation (NMF) and bilinear interpolation to estimate output spectra for arbitrary values of stellar age and metallicity.
The reduced disk requirements and computational expense allows the result to be served as a client-based Javascript application.
In this paper we present the method for generating grids of spectra, fitting those grids with NMF, bilinear interpolation across the fitted coefficients, and finally provide estimates of the prediction and interpolation errors.
\end{abstract}


\end{frontmatter}

% \linenumbers


\input{intro}
\input{methods}

\input{nmf}
\input{results}
\input{viewer}

\input{conc}



%% References


\section*{Acknowledgements}
We wish to thank the referees for comprehensive comments that significantly improved both the tool and this manuscript.
We also thank all of the SPS model owners for permission to include their models in Sengi.
Thanks also to Viviana Acquaviva and Jim Geach for useful comments and suggestions.
We acknowledge the following open source software packages used in the analysis, \textsc{scikit-learn} \citep{pedregosa_scikit-learn:_2011} and \textsc{Astropy} \citep{robitaille_astropy:_2013}.
Lovell acknowledges support from the Royal Society under grant RGF\textbackslash EA\textbackslash181016.

\bibliographystyle{elsarticle-harv}
\bibliography{sengi}






% \end{thebibliography}

\end{document}

%% file: intro.tex
\section{Introduction}

Stellar Population Synthesis (SPS) models predict the emission from stellar populations, typically coeval Simple Stellar Populations (SSP) with a single age, metallicity and abundance distribution.
To get from such a population to an SED requires stellar isochrones to describe stellar evolution, spectral libraries, and an Initial Mass Function (IMF).
The contribution of the effects of dust and nebular emission are subsequently added, with varying degrees of complexity, to produce realistic synthetic spectra.
A number of models have been developed by different groups over the past 20 years, and these are widely used across a range of astrophysical domains \citep[for a review, see][]{conroy_modeling_2013}.
Such models are still the subject of significant uncertainties, particularly for rarer phases of stellar evolution such as asymptotic giant branch (AGB) stars and blue stragglers.

There are, however, two draw backs to these codes for certain use-cases: their computational cost and large disk space requirements.
Downloading and analysing large libraries of models, or installing and running computationally expensive codes, is not an ideal or feasible approach in many use cases, particularly where computational resources are less accessible.
For example, in a prototyping or exploratory phase of a project these drawbacks can at best slow down the exploration of novel ideas, and at worst deter users from exploring them altogether.
Recent progress in a variety of fields has also highlighted that uncertainties in SPS models is a dominant source of uncertainty in inferred physical parameters \citep[\textit{e.g.}][]{wilkins_lyman-continuum_2016}.
The ability to quickly compare models would aid the qualitative evaluation of the impact of SPS model choice on a given analysis.

\textsc{Sengi} is an attempt to confront these issues by providing a small, fast, interactive web interface to some of the most well known SPS codes.
It does this using Non-negative Matrix Factorisation (NMF) applied to a grid of spectra generated from the models.
These grids are sampled for a range of ages and metallicities, two parameters that have a significant impact on the SSP emission.
Grids using different parameters, for example including the effects of nebular emission and dust attenuation, can also be generated.
The components and coefficients estimated through NMF can be interpolated to estimate the spectral emission for arbitrary parameters.
 % at the expense of relatively small interpolation errors.
\cite{kalmbach_estimating_2017} apply a similar technique to improving photometric redshift estimation, using gaussian process regression to perform the interpolation of coefficients derived using Principal Components Analysis (PCA).
We use bilinear interpolation, a computationally less expensive approach, which allows it to be estimated in a client-side javascript application.

Sengi is not a replacement for the original codes, since it cannot encapsulate the full flexibility in all parameters.
It is also not recommended for science pipelines; whilst we show in Sections \ref{sec:reconstruction} and \ref{sec:interpolation} that the recovery and interpolation errors are low, the original models should always be used for production science.
However, for prototyping and education \textsc{Sengi} provides a fast and accessible (low barrier-to-entry) tool.
\textsc{Sengi} is available online at \url{christopherlovell.github.io/sengi}, and the source code is made available at \url{https://github.com/christopherlovell/sengi} under the GNU General Public License v3.0.

This manuscript is arranged as follows.
In \sec{method} we describe our method, including the generation of grids (\ref{sec:generate_grids}), a description of the NMF algorithm used (\ref{sec:nmf}), our choice of hyperparameters (\ref{sec:hyperparameter}), an analysis of the fitted components (\ref{sec:components}), and a description of the interpolation method (\ref{sec:method_interp}).
Then in \sec{reconstruction} and \ref{sec:interpolation} we evaluate the reconstruction and interpolation accuracy, respectively.
Finally, we provide our conclusions and a discussion of the web-interface in \sec{discussion} and \ref{sec:conc}.

%% file: methods.tex
\section{Methods}
\label{sec:method}

% Below we describe our procedure for generating grids of outputs from SPS models.
% We then introduce our method for using these grids to fit a set of basis spectra and coefficients using non-negative matrix factorisation, and how this can be used to generate spectra for arbitrary parameters through linear interpolation.

\subsection{Generating grids of models}
\label{sec:generate_grids}

Our approach uses spectra generated from SPS models on a grid of age and metallicity values.
These grids are then fit using NMF.
There are a number of other factors, such as the IMF or the inclusion of nebular emission and dust, that have a significant effect on the emission.
Where these are discrete selections we can simply generate multiple grids, (\textit{e.g.} for different IMFs).
Where these choices are continuous (\textit{e.g.} varying extinction coefficient) we ignore them in this version, as this would enlarge the dimensionality, and subsequent disk size, of the grid substantially.

We have provided five grids in the current release, detailed in \tab{sps}.
These are constructed from three SPS models: \textsc{FSPS} \citep{conroy_propagation_2009,conroy_propagation_2010}, with nebular contributions included as calculated by \cite{byler_nebular_2017}, \textsc{BPASS} \citep{eldridge_binary_2017} and \textsc{BC03} \citep{bruzual_stellar_2003}.\footnote{We used python-FSPS \citep{dan_foreman-mackey_python-fsps:_2014} as a wrapper for FSPS to obtain the spectra, and the \textsc{hoki} interface \citep{stevance_hoki_2020} to read the BPASS spectra files. We obtained the BC03 models directly from \url{http://www.bruzual.org/bc03/Original_version_2003/}}
It is trivial to add new SPS models, or grids with parameter variations, simply by providing grids of spectra in age and metallicity.
The resolution of the grid should be sufficient to sample the variability in the spectra as a function of age and metallicity in order to reduce interpolation errors.
In the case of BC03 and BPASS we are limited by the outputs given in the data releases, however in FSPS we can generate arbitrarily fine grids.
We normalise each SSP to have mass $1 \, M_{\odot}$, and restrict the spectra to the UV - near infrared wavelength range ($[2 \times 10^{3}, 10^{4}]$ \si{\angstrom})
We assume a solar metallicity of $Z_{\odot} = 0.0127$ \citep{wiersma_chemical_2009}.

Further details on the models, including the age and metallicity resolution, are given in \tab{sps}.
We use the BPASS binary grid as our fiducial grid throughout the rest of the analysis, with a Chabrier IMF and a 300 $M_{\odot}$ high-mass cut off.

\begin{sidewaystable}
  \centering
  \begin{tabular}{l*{5}{c}r}
    Name & Model & IMF$^{1}$ & High-mass cutoff$^{2}$ & age [$\mathrm{log_{10}(Gyr)}$]$\;^{3}$ & Z [$\mathrm{log_{10}}(Z / Z_{\odot})$]$\;^{3}$ & $\mathrm{N_{spec}} \; ^4$ \\
    \hline
    FSPS & FSPS v3.1 & Kroupa & 120 & [-2,1.18,81] & [-2,1,41] & 3321 \\
    FSPS (Nebular) & FSPS v3.1 + Cloudy & Kroupa & 120 & [-2,1.18,81] & [-2,1,41] & 3321 \\
    BC03 & Galaxev (2003) & Chabrier & 100 & [-3.9,1.30,220] & [-4.84,1.37,6] & 1320 \\
    BPASS (Single) & BPASS v2.2.1 & Chabrier & 300 & [-3,1.2,43] & [-3.1,0.5,13] & 559 \\
    BPASS (Binary) & BPASS v2.2.1 & Chabrier & 300 & [-3,1.2,43] & [-3.1,0.5,13] & 559
    \end{tabular}
    \caption{SPS model grid details. (1) Initial Mass Function. (2) Upper limit to the Initial mass Function. (3) The lower limit, upper limit, and number of grid points (respectively) in each age and metallicity grid. (4) Number of spectra in each grid.}
  \label{tab:sps}
\end{sidewaystable}

%% file: nmf.tex
\subsection{Non-negative Matrix Factorisation}
\label{sec:nmf}

There are a number of dimensionality reduction techniques that use matrix factorisation of the form
\begin{equation}
  V_{ij} = (WH)_{ij} = \sum_{k=1}^{N} \, W_{ik} \, H_{kj} \;\;,
\end{equation}
where $V$ is some $n \times m$ matrix, with $n$ objects each composed of $m$ `features' in machine learning parlance.
The $N$ columns of $W$ represent the basis components, and each column of $H$ is an encoding for a given object.
Principal Component Analysis (PCA) is one of the most well known techniques of this form.
In PCA, the rows of $H$ are constrained to be orthogonal, and the columns of $W$ to be orthonormal.
This then leads to the property that the first component describes the direction along which maximum variance can be described.

Non-negative Matrix Factorisation (NMF) is another factorisation technique of this form.
NMF imposes the unique constraint that both $W$ and $H$ must be non-negative.
This is in contrast to PCA, where components can be negative, and the linear combination can be subtractive as well as additive.
The NMF approach has unique advantages for a number of physical applications, such as spectra decomposition, where negative luminosities are obviously un-physical \citep{hurley_learning_2014}.
This can also aid the interpretation of components \citep[see][]{lee_learning_1999}.

The initialised values of $W$ and $H$ can have a big impact on the subsequent performance of NMF.
We use Non-negative Double Singular Value Decomposition, where zeros are filled with small random values (known as NNDSVDar).
This initialisation approach leads to faster convergence \citep{boutsidis_svd_2008}.
The values of $W$ and $H$ are then updated using the multiplicative update (MU) algorithm, with the Itakura-Saito (IS) divergence of the two matrices as the objective function,
\begin{align}
  d_{IS}(X,Y) &=  \sum_{i,j} \left( \frac{X_{ij}}{Y_{ij}} - \mathrm{log}\frac{X_{ij}}{Y_{ij}} - 1 \right) \;\;,
\end{align}
where $X$ is the true grid spectra, and $Y = W \, H$ is the estimated grid spectra.
An alternative cost function is the Frobenius norm, which acts like a Euclidean distance for matrices.
IS works better for this application due to the large dynamical range of the grid spectra \citep{fevotte_nonnegative_2009}.
The update rules for MU with IS are \citep{abdallah_polyphonic_2004}:
\begin{align}
  H &\leftarrow H \frac{W^{T} \,((WH)^{-2} \, V)}{W^{T} \,(WH)^{-1}} \\
  W &\leftarrow W \frac{((WH)^{-2} \, V)\,H^{T}}{(WH)^{-1}\,H^{T}} \,\,.
\end{align}

\subsection{Hyperparameter optimisation}
\label{sec:hyperparameter}

A number of hyperparameters for NMF can be chosen to improve the prediction accuracy and performance time.
The tolerance, defined as the minimum reduction in the cost function between consecutive iterations, and the maximum number of iterations permitted to reach this tolerance, are two such hyperparameters.
We set the tolerance at $10^{-6}$, and the maximum number of iterations at $10^4$; this leads to convergence before the maximum iteration number in all cases shown here.

The number of components is another user-defined choice.
In PCA each component is unaffected by subsequent additional components.
In contrast, in NMF each additional component changes the existing components.
One could use the Bayesian evidence to rigorously determine the optimum component number \citep[see][]{hurley_learning_2014}, but since we are not chiefly concerned about the decomposed components in this application, we choose to select purely based on the trade off between predictive accuracy and optimum disk usage.

\begin{figure}
  \centering
\includegraphics[width=0.7\columnwidth]{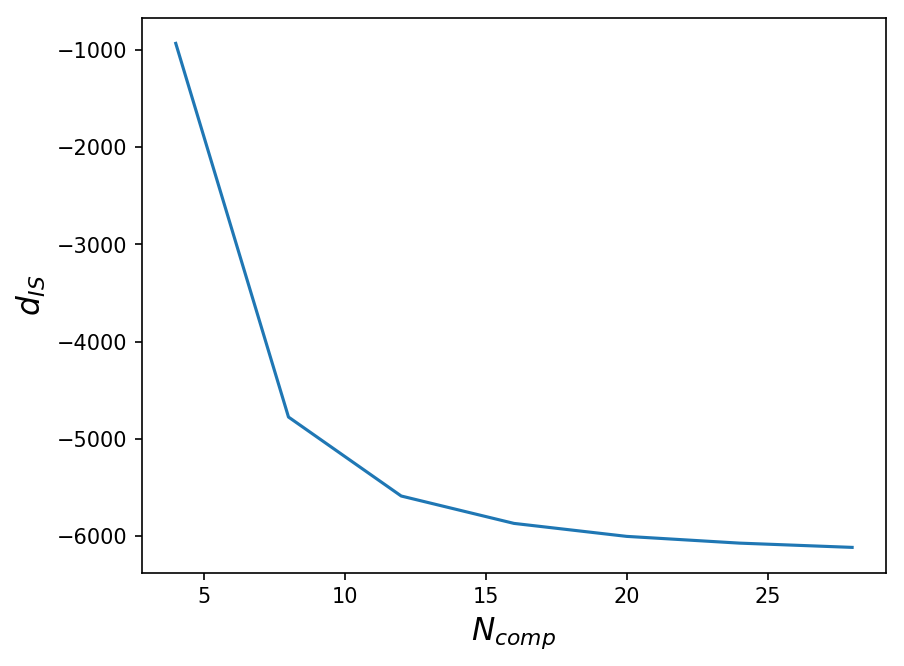}
    \caption{Itakura-Saito divergence, $d_{IS}$, against the number of components for the fiducial BPASS model (see Table \ref{tab:sps}).
    $d_{IS}$ plateaus at approximately $-6000$ for $>20$ components.}
    \label{fig:div_optim}
\end{figure}

\fig{div_optim} shows the average Itakura-Saito divergence, $d_{IS}$, against the number of components for the fiducial BPASS grid.
$d_{IS}$ is calculated between the true and the predicted spectra over the whole grid.
More components leads to a reduction in the error, but this plateau's at $\sim -6000$ for $>20$ components.

\begin{figure}
  \centering
\includegraphics[width=0.7\columnwidth]{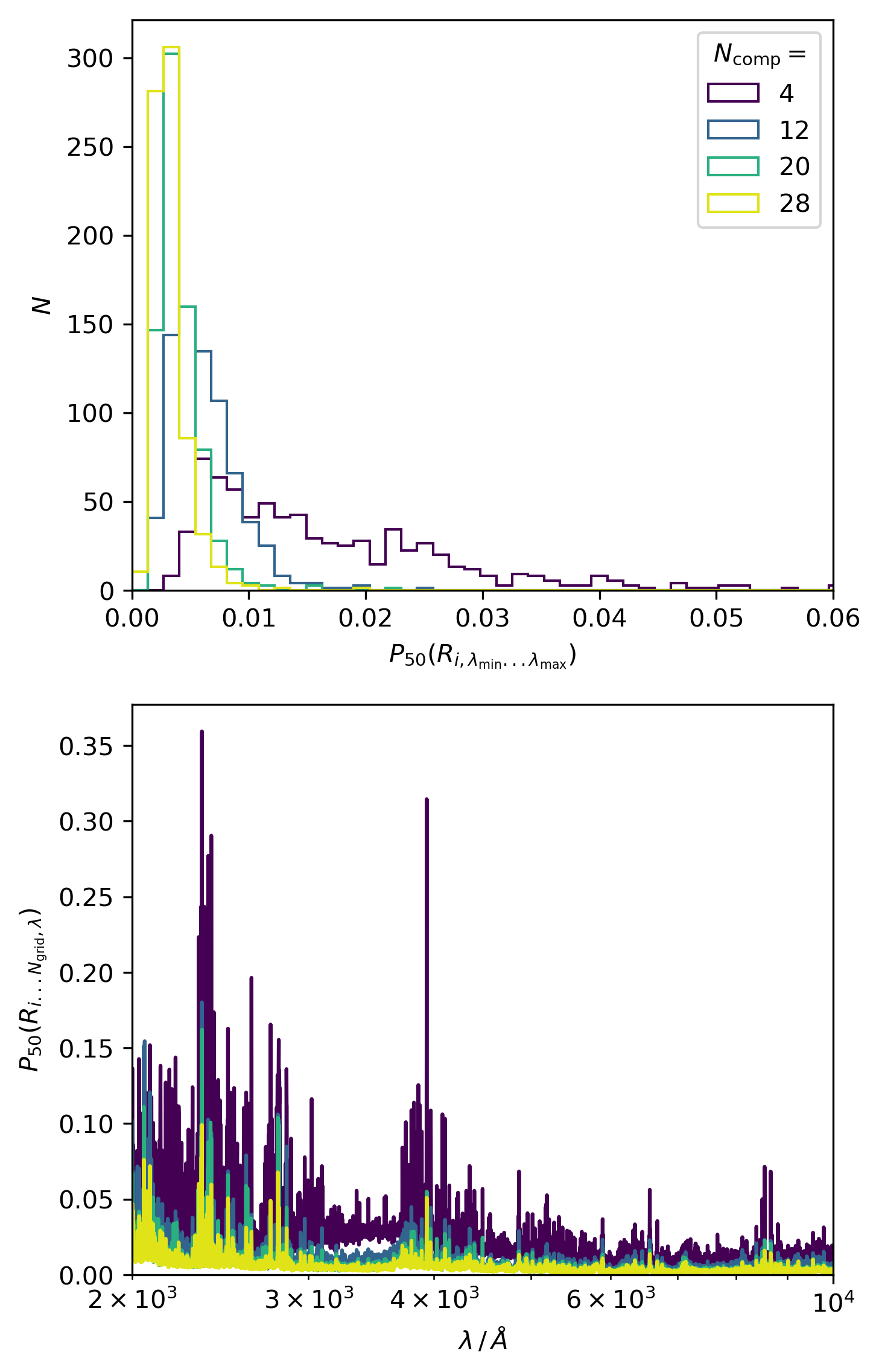}
    \caption{\textit{Top}: histogram of $P_{50} (R_{i,\lambda_{\mathrm{min}}...\lambda_{\mathrm{max}}})$, the median SMAPE for each grid point averaged across the whole spectrum, using the fiducial BPASS model. Each line shows a different number of components. \textit{Bottom}: $P_{50} (R_{i...N_{\mathrm{grid}},\lambda})$, the median SMAPE for all grid points ($N_{\mathrm{grid}}$) averaged over each pixel.}
    \label{fig:optim}
\end{figure}

We can also see the improvement in prediction accuracy by calculating the Symmetric Mean Absolute Percentage Error (SMAPE), defined as
\begin{align}
  R_{i,\lambda} = 2 \times \frac{\left| S^{\mathrm{pred}}_{i,\lambda} - S^{\mathrm{true}}_{i,\lambda} \right|}{S^{\mathrm{pred}}_{i,\lambda} + S^{\mathrm{true}}_{i,\lambda}} \;\;,
\end{align}
where $R$ is the SMAPE for grid point $i$, wavelength $\lambda$, and $S^{\mathrm{pred}}_{i,\lambda}$ and $S^{\mathrm{true}}_{i,\lambda}$ are the predicted and true spectra, respectively.
We use SMAPE as it is less sensitive to large changes in dynamic range \citep[see][]{lovell_learning_2019}, giving a fairer comparison of the model residual for different grid spectra.
SMAPE acts as a true percentage error in the small error regime.

The top panel of \fig{optim} shows the distribution of the median of $R$, $P_{50} (R_{i,\lambda_{\mathrm{min}}...\lambda_{\mathrm{max}}})$, for all grid points $i$.
For $N_{\mathrm{comp}} = 20$ the error is below 3.1\% for all grid points, with a median at $\sim$0.4\%.
The bottom panel of \fig{optim} shows $P_{50} (R_{i...N_{\mathrm{grid}},\lambda})$ for each wavelength.
For $N_{\mathrm{comp}} = 20$, it peaks at $\sim$16\% for some emission lines, but the average across the spectrum is still in the single-digit percent regime.

We choose $N_{\mathrm{comp}} = 20$, as a trade-off between accuracy of the prediction and disk size of the grid.

\subsection{Components}
\label{sec:components}

\begin{figure*}
\includegraphics[width=\textwidth]{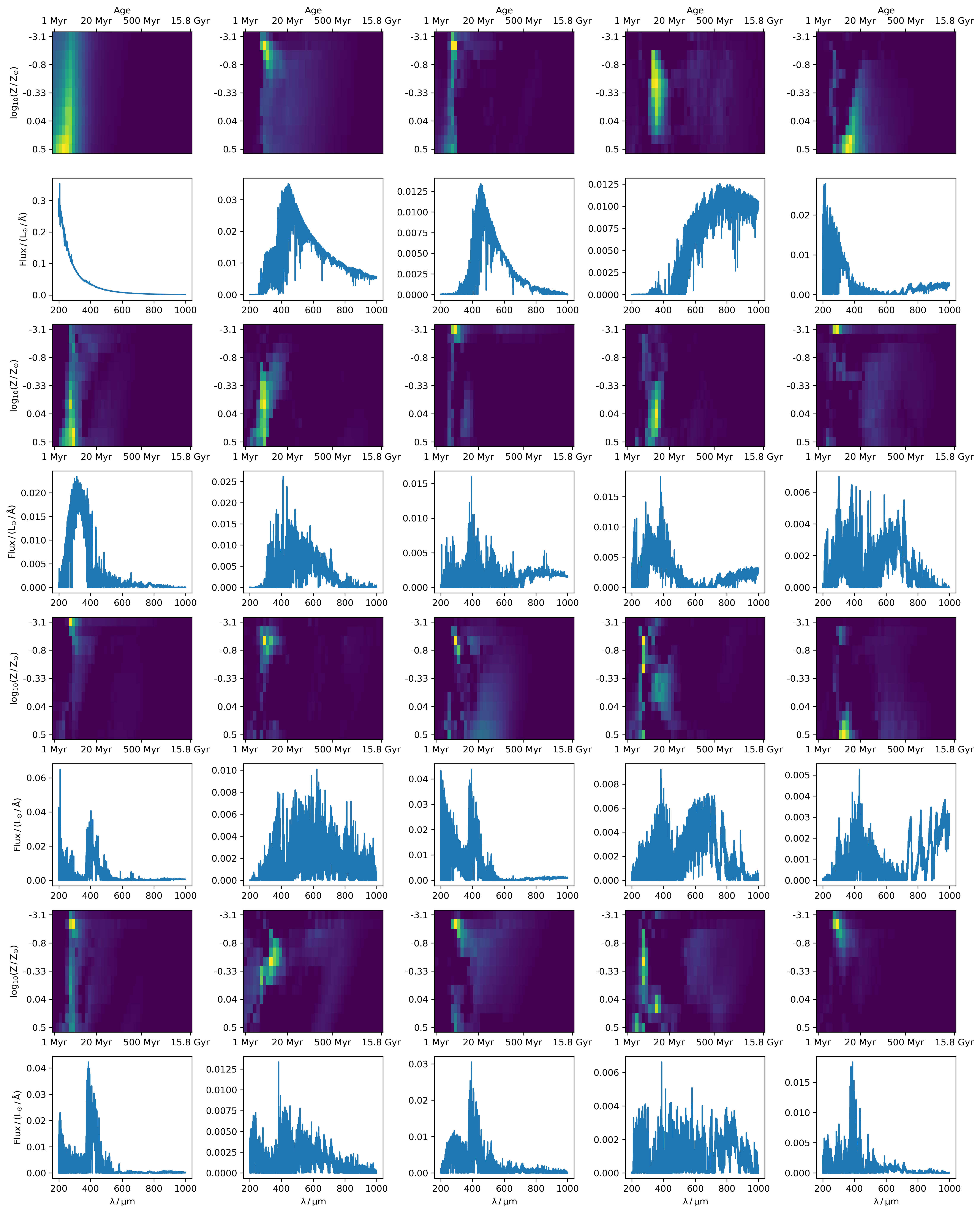}
    \caption{Components and coefficient weights matrices for the fiducial \textsc{BPASS} grid. Each grid plot shows the matrix of coefficient weights, and directly below is shown the component spectra. Components are shown from 1-20, labelled on the coefficient matrix.}
    \label{fig:coeffs_nmf}
\end{figure*}

\fig{coeffs_nmf} shows each component and its matrix coefficients for the fiducial BPASS grid.
The first component dominates in flux over all other components, and is associated with young, high metallicity populations.
Subsequent components sample the rest of the age - metallicity plane.

\subsection{Grid Interpolation}
\label{sec:method_interp}

The coefficients and components estimated through NMF can be used to predict the spectra for each isolated grid point.
Additionally, using appropriate interpolation of those coefficients across the grid, an estimate of the spectra can be obtained for arbitrary grid parameters (within the grid bounds).
We use bilinear interpolation, which gives a good trade-off between interpolation accuracy and computational cost.
The latter constraint is important, as the interpolation is performed on the client side in the \textsc{Sengi} web application.
The accuracy of the interpolation is assessed in \sec{interpolation}.

%% file: results.tex
\section{Reconstruction Error}
\label{sec:reconstruction}

\begin{figure}
\includegraphics[width=\columnwidth]{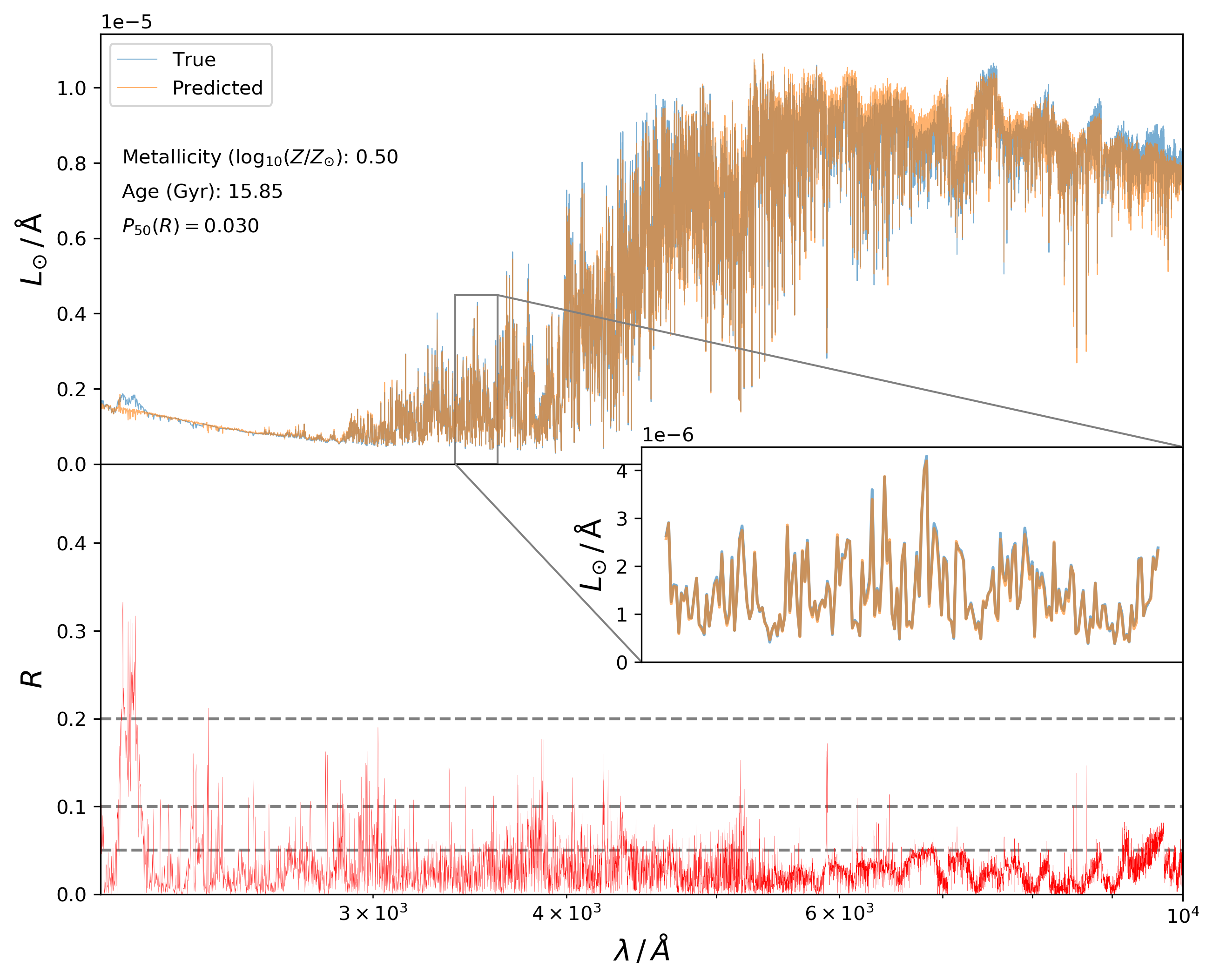}
    \caption{\textit{Top:} example predicted spectra (orange) compared to the true spectra (blue) for the grid point with the \textbf{highest error} in the fiducial BPASS grid ($R = 0.03$).
    \textit{Inset:} zoom on a region of the spectrum, showing the good agreement of the detailed spectral features.
    \textit{Bottom:} $R$ against wavelength.}
    \label{fig:errs_example_nmf}
\end{figure}

We first evaluate the prediction accuracy on the grid.
An example prediction for the grid point with the \textbf{highest median error}, $R = 0.03$ (see \fig{optim}), is shown in \fig{errs_example_nmf}.
The predicted spectrum recovers both the overall shape and high resolution details of the spectrum remarkably well.
$R$ is consistently below 2\% across the wavelength range, except for a few pixels at the blue end of the spectrum, where NMF struggles to reproduce a feature in the spectrum.
The errors are consistently at the sub-percent level at the red end of the spectrum.

\begin{figure}
  \centering
\includegraphics[width=0.7\columnwidth]{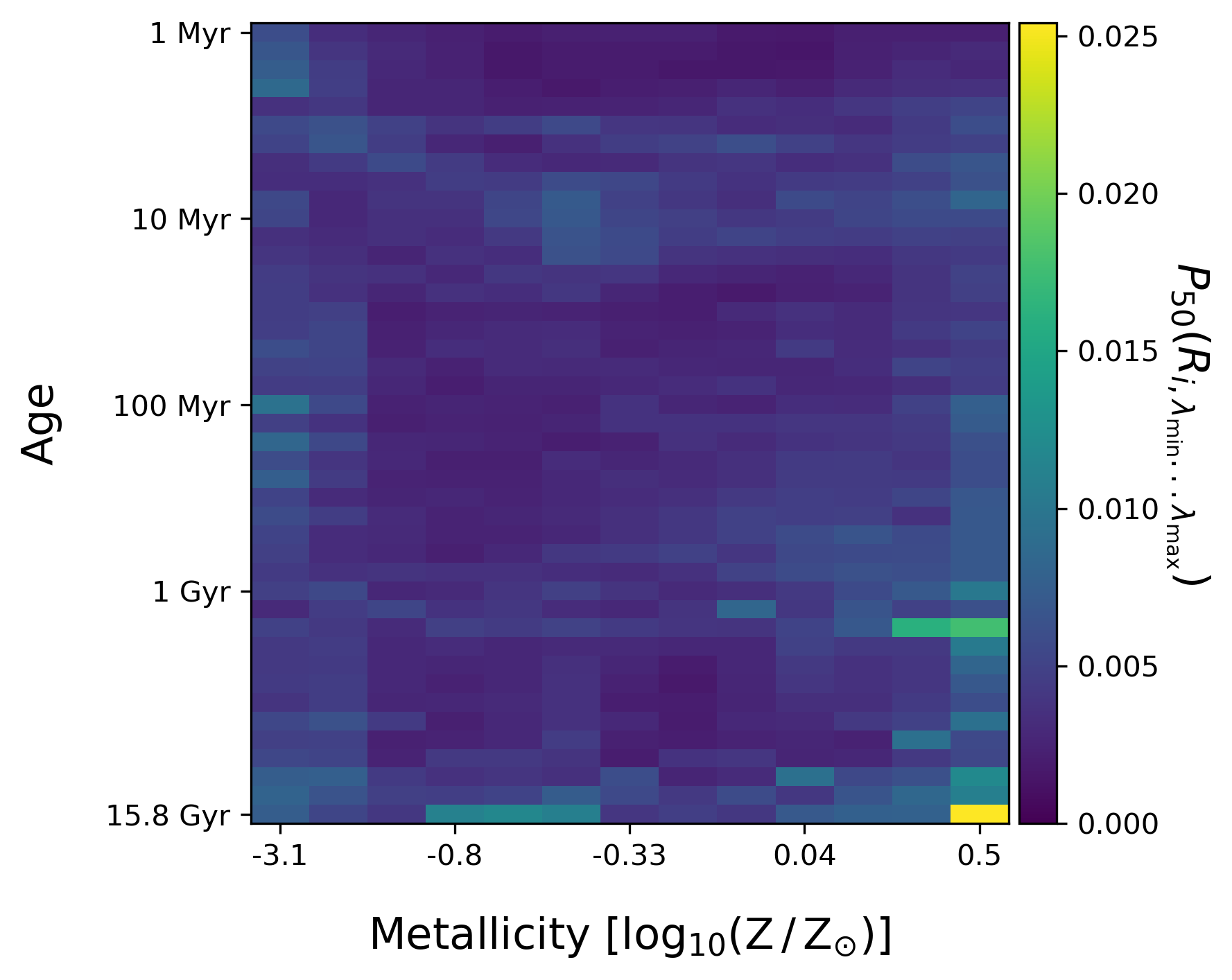}
    \caption{Median error, $P_{50} (R_{i,\lambda_{\mathrm{min}}...\lambda_{\mathrm{max}}})$, between the predicted and true spectra across the age - metallicity grid, for the fiducial BPASS model. Reconstruction errors tend to be larger at the grid edges.}
    \label{fig:errmat_nmf}
\end{figure}

\fig{errmat_nmf} shows the median error across the spectrum for each grid point $i$, $P_{50} (R_{i,\lambda_{\mathrm{min}}...\lambda_{\mathrm{max}}})$.
Over the entire grid the median error is $< 1$\%.
$R$ tends to be higher at the edges of the grid, as expected since there are fewer neighbouring spectra in age - metallicity space for components to be built from.
In the centre of the grid, where there is a better sampling of similar spectra, the errors tend to be lower.

\section{Interpolation Error}
\label{sec:interpolation}

NMF provides the coefficient weights for all points on the grid, which can be used to predict the emission for discrete grid points, as shown in \sec{reconstruction}.
These coefficients can also be used to estimate the emission for arbitrary values of age and metallicity (within the grid bounds) through interpolation of each coefficient.
We use bilinear interpolation due to its speed and low computational complexity.
\cite{kalmbach_estimating_2017} show that using more sophisticated methods, such as 2D gaussian process regression, can reduce the interpolation error.
However, such techniques are currently not suitable for client-side web applications due to the increased computational cost.

\begin{figure}
  \centering
  \includegraphics[width=0.7\columnwidth]{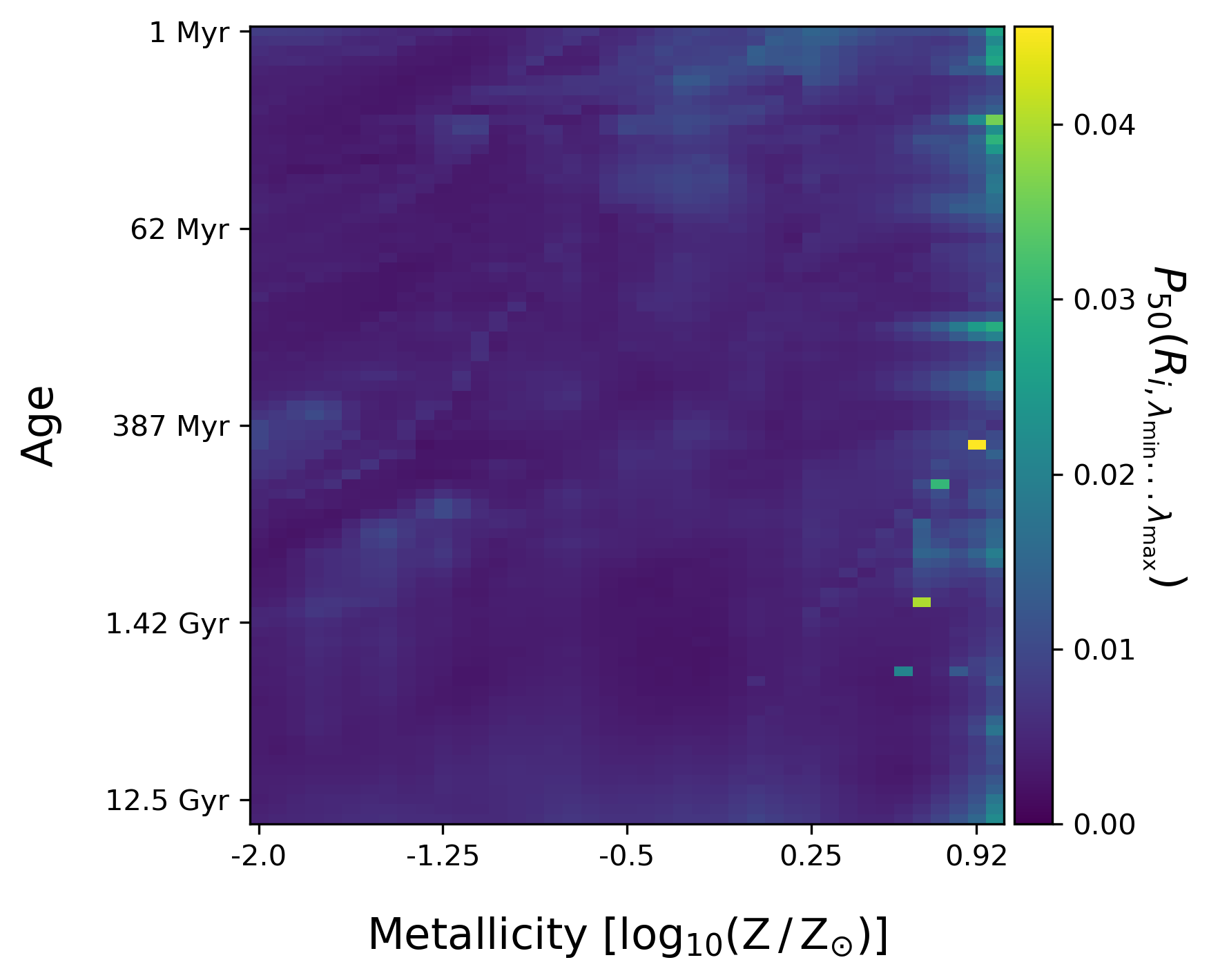}
\includegraphics[width=0.7\columnwidth]{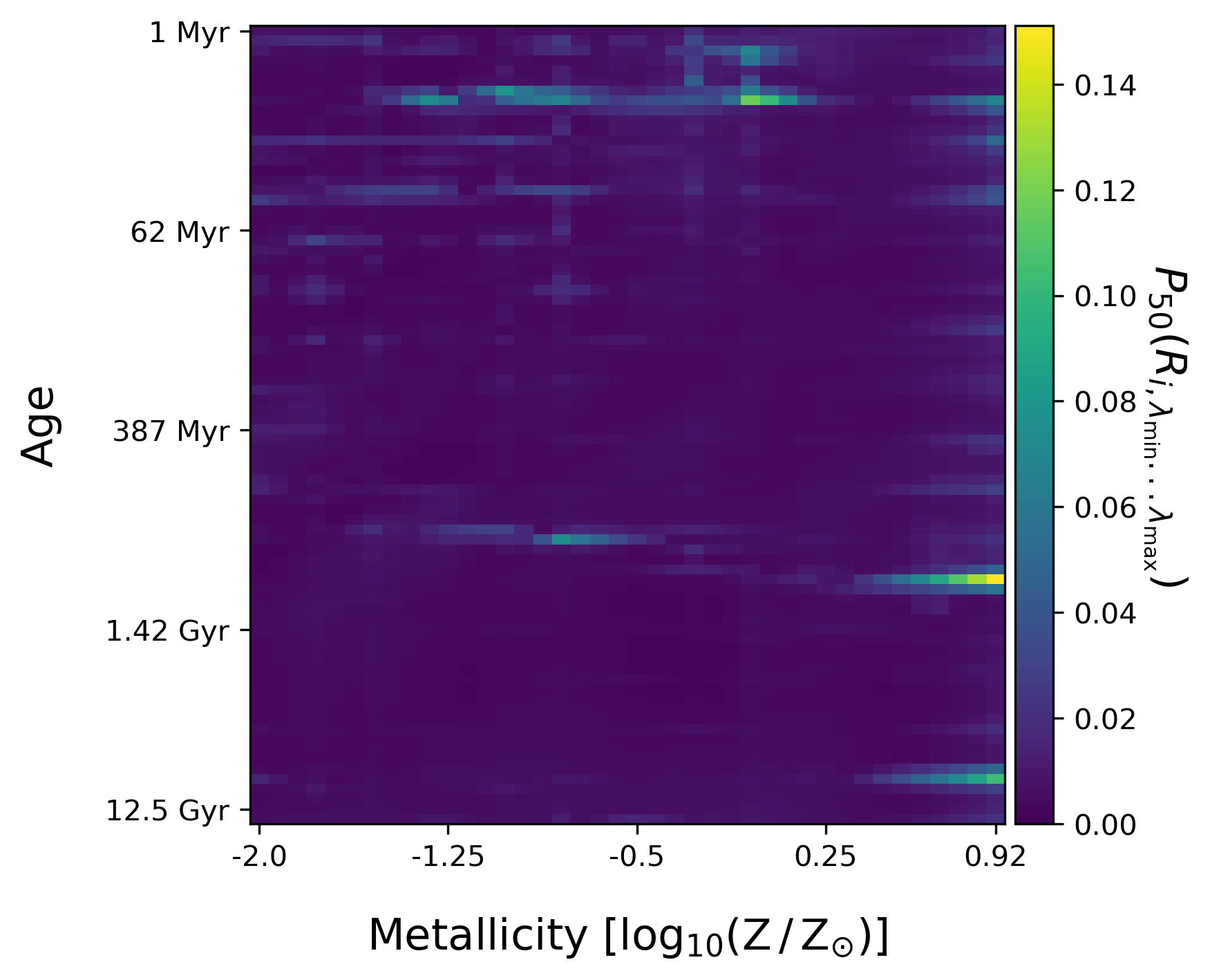}
    \label{fig:interp_errmat}
    \caption{Median error, $P_{50} (R_{i,\lambda_{\mathrm{min}}...\lambda_{\mathrm{max}}})$, across the age - metallicity grid for the FSPS model. \textit{Top:} error between the grid and the predicted spectra. \textit{Bottom:} error between the high resolution grid and the interpolated predicted spectra.}
\end{figure}

We evaluate the error introduced through interpolation by generating two grids, one with twice the resolution in age and metallicity.
We use python-FSPS to generate the grids as it allows us to sample the age and metallicity at arbitrary resolution.
For the high resolution grid we produce spectra for 161 age values, sampled regularly between -2 and 1.17 in $\mathrm{log_{10}(Gyr)}$, and 81 metallicity values sampled regularly between -2 and -1 in $\mathrm{log_{10}(Z\,/\, Z_{\odot})}$.
For the lower resolution grid we use the same limits but with half the number of grid points.

This leaves 3200 points on the high resolution grid within bounds of the lower resolution grid.
We then estimate the values on these points by running NMF on the lower resolution grid, interpolating the coefficients at each high-resolution grid point.
The bottom panel of \fig{interp_errmat} shows the error between the predicted spectra on these un-sampled points and the true spectra from the high resolution grid.
The mean of $R$ across the grid is 0.75\%, and the median is 0.51\%.
This is comparable to the reconstruction errors on the grid points, shown in the top panel of \fig{interp_errmat} (analogous to \fig{errmat_nmf} but for FSPS).
However, there are peaks in the distribution of $R$, particularly at the edges of the grid, where interpolation errors can reach $\sim \, 15\%$.

Using python-FSPS we can mitigate the effect of interpolation errors at the grid edges by creating both wider and finer grids, but this is not currently possible for all SPS models.

%% file: viewer.tex
% \section{Sengi viewer}
%
% \begin{figure*}
%     \centering
%     \includegraphics[width=\textwidth]{images/screenshot.png}
%     \caption{The \textsc{Sengi} viewer.}
%     \label{fig:screenshot}
% \end{figure*}
%
% \fig{screenshot} shows a screenshot of the Sengi viewer.
% The plot at the centre of the page shows the predicted spectra for two different models.
% Additional models can be viewed by clicking the `Add line' button.
% The mouse can be used to zoom on a particular wavelength region of the spectrum by clicking and dragging over the plotting area.
% Selections can be cancelled by double clicking the plotting area.
% Each line has a set of controls at the bottom of page, labelled with the chosen model, age and metallicity, and distinguishable by a circle with the same colour as the line.
% Each set of controls can be activated by clicking on the header.
% The age and metallicity can be adjusted using the sliders.
% New models can be chosen using the drop-down selector.

%% file: conc.tex
\section{Discussion}
\label{sec:discussion}

Whilst \textsc{Sengi} has been designed as an interactive online tool, the method of spectra reconstruction presented here can be used for a wide range of applications.
\cite{kalmbach_estimating_2017} first showed that using SEDs estimated from a set of basis components can lead to much higher precision photometric redshifts.
Recently, coincident with submission of this manuscript, \cite{alsing_speculator:_2020} showed how using neural networks combined with basis components derived from PCA can be used to flexibly estimate coefficients for arbitrary SPS parameter coefficients.
They use this within an SED fitting framework to estimate physical parameters of galaxies, and find significant computational speed ups over generating spectra directly from the models, as we find here.
We note that they apply PCA to spectra in \textit{log} flux; in our tests we found much improved performance when evaluated in this space, due to the smaller dynamic range.
However, when the fitted spectra were evaluated in linear flux the errors were much larger.
The quoted fractional errors in log flux in \cite{alsing_speculator:_2020} are therefore not directly comparable with the SMAPE errors calculated here from the linear flux.
The neural network developed in \cite{alsing_speculator:_2020} is written in Tensorflow; the trained model could be translated to a web framework through \texttt{tensorflow.js}.\footnote{\url{https://www.tensorflow.org/js}}

Cosmological simulations are another area where spectra estimation could provide useful speed ups and reductions in data footprint.
`Forward modelling' simulations is a key method of testing their predictions directly with observables.
However, this is typically done for at most a limited number of SED modelling assumptions, and the full SEDs are usually not provided.
Instead, photometry is usually the main data product due to the lower data cost \citep{camps_data_2018,torrey_synthetic_2015}.
Spectra estimation from basis components would allow for the full modelled spectra from a large set of assumed parameters to be provided given a relatively small data footprint.
Spectra generation in simulations also typically uses SSP spectra generated over a grid of age and metallicity that are then interpolated to give the emission for arbitrary stellar elements.
Basis estimation could allow more rapid, and potentially more accurate, generation of spectra from the simulations for a larger range of modelling assumptions.
This capability will be key to understanding the degeneracies and biases imposed by these assumptions \citep{wilkins_lyman-continuum_2016}.

Sengi has a number of properties that make it ideal for use as an educational tool.
The web tool is freely accessible online via a browser, and is therefore platform independent, requiring no setup from the user.
For astronomy educators this significantly reduces the barrier to entry compared to using the SPS model data directly, and the local computational storage and processing required.
The almost instantaneous updating of the plotted spectra when models are changed or added allows the tool to be used interactively.
As an example, Sengi could be used in an undergraduate stellar evolution class, to demonstrate the effect of age and metallicity on stellar emission during a lecture.
It could also be used in more advanced courses to highlight the impact of SPS model choice on the predicted emission at fixed age and metallicity.

\section{Conclusions}
\label{sec:conc}

We have presented a novel method for representing the spectral results from SPS models, using Non-negative Matrix Factorisation and bilinear interpolation.
This bypasses the computational complexity of generating the models, and reduces the data footprint, allowing the method to be presented via a client-side Javascript application.
\textsc{Sengi} is available online at \url{christopherlovell.github.io/sengi}, providing a spectral viewer for each of the five grids described in this paper.
Interactive sliders allow users to change the age and metallicity of the SSP and instantly see the effect on the optical emission.
Multiple models can be viewed simultaneously, and users can zoom in to areas of the spectra of interest.

The source code is available online at \url{https://github.com/christopherlovell/sengi} under the GNU General Public License v3.0.